# Fractional quantum conductance staircase of edge hole channels in silicon quantum wells

N.T. Bagraev*, L.E. Klyachkin, A.A. Kudryavtsev, and A.M. Malyarenko

*Ioffe Physical-Technical Institute, 194021 St. Petersburg, Russia*

**Abstract**

We present the findings for the fractional quantum conductance of holes that is caused by the edge channels in the silicon nanosandwich prepared within frameworks of the Hall geometry. This nanosandwich represents the ultra-narrow p-type silicon quantum well (Si-QW), 2 nm, confined by the δ-barriers heavily doped with boron on the n-type Si (100) surface. The edge channels in the Si-QW plane are revealed by measuring the longitudinal quantum conductance staircase, $G_{xx}$, as a function of the voltage applied to the Hall contacts, $V_{xy}$, to a maximum of $4e^2/h$. In addition to the standard plateau, $2e^2/h$, the variations of the $V_{xy}$ voltage appear to exhibit the fractional form of the quantum conductance staircase with the plateaus and steps that bring into correlation respectively with the odd and even fractional values.

*Introduction* – At present, the methods of a semiconductor nanotechnology such as the split-gate [1-3] and cleaved edge overgrowth [4] allow the fabrication of the quasi-one-dimensional (1D) constrictions with low density high mobility carriers, which exhibit the characteristics of ballistic transport. The conductance of such quantum wires with the length shorter than the mean free path is quantized in units of $G_0 = g_s e^2/h$ depending on the number of the occupied 1D channels, $N$. This quantum conductance staircase, $G = G_0 \cdot N$, has been revealed by varying the split-gate voltage applied to the electron and hole GaAs-[1-7] and Si-based [8-10] quantum wires. It is significant that spin factor, $g_s$, that describes the spin degeneration of carriers in a 1D channel appears to be equal to 2 for noninteracting fermions if the external magnetic field is absent and becomes unity as a result of the Zeeman splitting of a quantum conductance staircase in strong magnetic field.

However, the first step of the quantum conductance staircase has been found to split off near the value of $0.7(2e^2/h)$ in a zero magnetic field. Two experimental observations indicate the importance of the spin component for the behavior of this $0.7(2e^2/h)$ feature. Firstly, the electron *g* factor was measured to increase from 0.4 to 1.3 as the number of occupied 1D subbands decreases [5]. Secondly, the height of the $0.7(2e^2/h)$ feature attains a value of $0.5(2e^2/h)$ with increasing external magnetic field [5,11,12]. These results have defined the spontaneous spin polarization of a 1D gas in a zero magnetic field as one of possible mechanisms for the $0.7(2e^2/h)$ feature in spite of the theoretical prediction of a ferromagnetic state instability in ideal 1D system in the absence of magnetic field [13].

More recently, the $0.7(2e^2/h)$ feature has been shown to be close also to the value of $0.5(2e^2/h)$ at low sheet density of both holes [9,10] and electrons [14]. These measurements performed by tuning the top gate voltage that allows the sheet density control have revealed not only the principal role of the spontaneous spin polarization, but the changes of the $0.7(2e^2/h)$ feature in fractional form as well [10]. Therefore the studies of the $0.7(2e^2/h)$ feature are challenging to be explored in the topological insulators or superconductors in which the one-dimensional, 1D, edge channels are formed without any electrical or mechanical restriction [15,16]. These edge channels are similarity to the quantum Hall states wherein the electric current is carried only along the edge of the sample, but exhibit helical properties, because two carriers with opposite spin polarization counter propagate at a given edge [17].

However, the quantum conductance staircase caused by the carriers in the edge channels is still uninvestigated in a zero magnetic field. Here we demonstrate the quantum conductance staircase of holes that is revealed by varying the voltage applied to the Hall contacts prepared at edges of the ultra-narrow p-type silicon quantum well, Si-QW. The fractional features that emerge in this conductance staircase appear to be evidence of the high spin polarization of holes in the helical edge channels. The latter findings were made possible by the developments of the diffusion nanotechnology that allows the fabrication of the nanosandwiches prepared on the n-type Si (100) surface as the ultra-narrow p-type Si-QW, 2 nm, confined by the δ -barriers heavily doped with boron (see Fig. 1) [8-10, 18].

*Device* – The silicon nanosandwich was prepared within frameworks of the Hall geometry after preliminary oxidation of the n-type Si (100) wafer, making a mask, performing the photolithography and subsequent short-time diffusion of boron by the CVD-related method [8,18]. The energy positions of the subbands for the 2D heavy, HH1: $E_V$ – 90 meV, and light, LH1: $E_V$ – 114 meV, holes in the Si-QW were determined by studying the far-infrared electroluminescence spectra with the infrared Fourier spectrometer IFS-115 Brucker Physik AG as well as by using the local tunneling spectroscopy technique [10,18]. The results obtained are in a good agreement with corresponding calculations following by Ref [19] if the width of the Si-QW, 2nm, is taken into account.

The secondary ion mass spectroscopy (SIMS) and scanning tunneling microscopy (STM) studies have shown that the δ - barriers, 3 nm, heavily doped with boron, $5 \cdot 10^{21}$ cm$^{-3}$, represent really alternating arrays of undoped and doped tetrahedral dots with dimensions restricted to 2 nm. The concentration of boron determined by the SIMS method seems to indicate that each doped dot located between undoped dots contains two impurity atoms of boron. Nevertheless, the angular dependencies of the cyclotron resonance spectra and the conductivity have demonstrated that these p-type Si-QW confined by the δ - barriers heavily doped with boron contain the high mobility 2D hole gas that is characterized by long transport relaxation time of heavy and light holes at 3.8 K, $\tau \geq 5 \cdot 10^{-10}$ *c* [8, 10]. Thus, the transport relaxation time of holes in the ultra-narrow Si-QW appeared to be longer than in the best MOS structures contrary to what might be expected from strong scattering by the heavily doped δ - barriers. This passive role of the δ - barriers between which the Si-QW is formed was quite surprising, when one takes into account the high level of their boron doping. To eliminate this problem, the temperature dependencies of the conductivity and the Seebeck coefficient as well as the EPR spectra and the local tunneling current-voltage characteristics have been studied [18]. The EPR and the thermo-emf studies have shown that the boron pairs inside the δ - barriers are the trigonal dipole centres, $B^+$-$B^-$, which are caused by the negative-U reconstruction of the shallow boron acceptors, $2B_0 => B^+ + B^-$. These extraordinary properties of the δ – barriers heavily doped with boron appeared to give rise to the small effective mass of the heavy holes, $<6·10^{-4}$ $m_0$, which was controlled by measuring the Aharonov – Casher conductance oscillations and the temperature dependences of the Shubnikov – de Haas (SdH) oscillations [10]. Therefore the Hall measurements at the temperature of 3.8K of the sheet density and the mobility in the experimental sample, $p_{2D}=9·10^{13}$ $m^{-2}$, $\mu=420$ $m^2V^{-1}s^{-1}$, conform to the high mobility low density gas of 2D holes. Besides, the value of the mobility showed a decrease no more than two times in the range from 3.8 to 77 K that appears to be in agreement with the temperature dependence of the transport relaxation time [18]. These characteristics of the 2D gas of holes in the silicon nanosandwiches made it possible for the first time to use the split-gate constriction to study the $0.7(2e^2/h)$ feature in the quantum conductance staircase of holes at the temperature of 77 K [8-10]. Furthermore, even with small drain-source voltage the electrostatically ordered dipole centres of boron within the δ - barriers appeared to stabilize the formation of the one-dimensional subbands, when the quantum wires are created inside Si-QW using the split-gate technique [8]. Thus, the ultra-narrow p-type Si-QW confined by the δ - barriers heavily doped with boron in properties and composition seems to be similar to graphene [20] that enables one to observe the fractional quantum conductance staircase of holes by varying the voltage applied to the Hall contacts which is able to control the changes of the spin polarization in the edge channels (see Fig. 1).

*Results* - The $R_{xx}$ dependence on the bias voltage applied to the Hall contacts, $V_{xy}$, exhibits the quantum conductance staircase to a maximum of $4e^2/h$ (Fig. 2a). This conductance feature appeared to be independent of the sample geometric parameters that should point out on the formation of the edge channels in Si-QW. Therefore we assume that the maximum number of these channels is equal to 2, one for up-spin and other for down-spin. It should be noted that the important condition to register this quantum conductance staircase is to stabilize the drain-source current at the value of lower than 1 nA.

In addition to the standard plateau, $2e^2/h$, the quantum conductance staircase appears to reveal the distinguishing features as the plateaus and steps that bring into correlation respectively with the odd and even fractions. Since similar quantum conductance staircase was observed by varying the top gate voltage which controls the sheet density of carriers and thus can be favorable to the spontaneous spin polarization, the variations of the $V_{xy}$ value seem also to result in the same effect on the longitudinal resistance, $R_{xx}$ (Figs. 2b and c).

The $R_{xx}$ fractional values revealed by tuning the $V_{xy}$ voltage appear to evidence that the only closely adjacent helical channels to the edge of Si-QW make dominating contribution in the quantum conductance staircase as distinguished from the internal channels. The device used implies a vertical position of helical channels, with quantum point contact (QPC) inserted as a result of the local disorder in the δ–barriers heavily doped with boron (Fig. 3). Besides, if the silicon nanosandwich is taken into account to be prepared along the [011] axis (Fig. 1), the trigonal dipole boron centers ordered similarly appear to give rise to the formation of the helical edge channels in Si-QW. It should be noted that depending on the degree of disorder in the δ–barriers the helical channels have to reveal the insulating or superconducting properties thereby defining the regime of the spin-dependent transport through the ballistic QPC [21]. In the latter case the multiple Andreev reflections seem to result in the spin polarization of carriers in helical channels in addition to the mechanism of spontaneous spin polarization [22,23].

The exchange interaction between holes localized and propagating through the QPC inside a quantum wire has been shown to give rise to the fractional quantization of the conductance unlike that in electronic systems [24]. Both the offset between the bands of the heavy and light holes, Δ, and the sign of the exchange interaction constant appeared to affect on the observed value of the conductance at the additional plateaus. Within the framework of this approach for the Si-based quantum wire, the conductance plateaus have to be close to the values of $e^2/4h$, $e^2/h$ and $9e^2/4h$ from the predominant antiferromagnetic interaction, whereas the prevailing ferromagnetic exchange interaction has to result in the plateaus at the values of $7e^2/4h$, $3e^2/h$ and $15e^2/4h$ [24]. Bearing in mind this prediction, we have observed the fractional features, which are in a good agreement with the ferromagnetic interaction theory (Figs. 2a, b and c). But the reason why the even fractional values correspond to the middle of the steps in the quantum conductance staircase instead of the plateaus as predicted in Ref [24] is needed to be analyzed in detail. Since the odd and even fractions have been exhibited similarly in the quantum conductance staircase as a function of the top gate voltage that controls the sheet density of holes, the parallels between the helical channels and the ballistic channels responsible for the fractional quantum Hall effect in strong magnetic fields engage once again our attention.

As mentioned above, the helical edge channels could be used to verify the interplay between the amplitude of the $0.7(2e^2/h)$ feature and the degree of the spontaneous spin polarization when the energy of the ferromagnetic exchange interaction begins to exceed the kinetic energy in a zero magnetic field. Specifically, the evolution of the $0.7(2e^2/h)$ feature in the quantum conductance staircase from $e^2/h$ to $3/2(e^2/h)$ is shown in Fig. 2c, which seems to be caused by the spin depolarization processes in QPC [9,25,26]. It is appropriate to suggest that the spin depolarization is the basic mechanism of the quantum conductance staircase as a function of the $V_{xy}$ bias voltage. This consideration can be best be done within framework of the Landauer-Buttiker formalism [27,28], if the $V_{xy}$ bias voltage is used to backscatter helical edge channels only on the one side of the device (Fig. 4). As a result of this backscattering, that is different in adjacent channels, in which the carriers counter propagate, the total spin polarization has to be varied thus providing the quantum conductance staircase as a function of the $V_{xy}$ bias voltage. Experimentally, the high resolved variations of the $V_{xy}$ bias voltage appeared to cause the observation of the exotic plateaus and steps (see the inset in Fig. 2a) that are evidence of the hole particles with fractional statistics [29], which are also needed to be studied in detail.

In summary, we have found the fractional form of the longitudinal quantum conductance staircase of holes, $G_{xx}$, that was measured as a function of the $V_{xy}$ bias voltage in the p-type silicon quantum well confined by the δ-barriers heavily doped with boron, which is prepared in the framework of the Hall geometry. The fractional conductance features observed at the values of $7e^2/4h$, $3e^2/h$ and $15e^2/4h$ are evidence of the prevailing ferromagnetic exchange interaction that gives rise to the spin polarization of holes in a zero magnetic field. This quantum conductance staircase measured to a maximum of $4e^2/h$, with the plateaus and steps that bring into correlation respectively with the odd and even fractional values, seems to reveal the formation of the helical edge channels in the p-type silicon quantum well.

[1] T. J. Thornton, M. Pepper, H. Ahmed, D. Andrews, and G. J. Davies, Phys. Rev. Lett. **56**, 1198 (1986).
[2] D.A. Wharam, *et al.*, J. Phys. C **21**, L209 (1988).
[3] B. J. van Wees, *et al.*, Phys. Rev. Lett. **60**, 848 (1988).
[4] A. Yacoby, *et al.*, Phys. Rev. Lett. **77**, 4612 (1996).
[5] K. J. Thomas, *et al.*, Phys. Rev. Lett. **77**, 135 (1996).
[6] L. P. Rokhinson, L. N. Pfeiffer, and K. W. West, Phys. Rev. Lett. **96**, 156602 (2006).
[7] O. Klochan, *et al.,* Appl. Phys. Lett. **89**, 092105 (2006).
[8] N. T. Bagraev, *et al.*, Semiconductors **36,** 439 (2002).
[9] N. T. Bagraev, *et al.*, Phys. Rev. B **70,** 155315 (2004).
[10] N. T. Bagraev, et al., J. Phys.: Condens. Matter, 20 164202, (2008).
[11] K. J. Thomas, *et al.*, Phys. Rev., B **58**, 4846 (1998).
[12] K. J. Thomas, *et al.*, Phys. Rev., B **61**, 13365 (2000).

[13] E. Lieb and D. Mattis, Phys. Rev., **125**, 164 (1962).

[14] R. Crook, *et al.*, Science **312**, 1359 (2006).

[15] M. Z. Hasan and C. L. Kane, Rev. Mod. Phys. **82**, 3045 (2010).

[16] Xiao-Liang Qi and Shou-Cheng Zhang, Rev. Mod. Phys. **83**, 1057 (2011).

[17] M. Buttiker, Science **325**, 278 (2009).

[18] N. T. Bagraev, *et al.*,: Superconductor Properties for Silicon Nanostructures. In "Superconductivity - Theory and Applications", edited by A. Luiz, SCIYO, 2010, chap.4, p.p.69-92.

[19] C. Weisbuch and B. Winter, Quantum Semiconductor Structures, p.15, Academic Press, New York, (1991).

[20] A. K. Geim and K. S. Novoselov, Nature Materials **6,** 183 (2007).

[21] N. T. Bagraev, *et al.*, Semiconductors **46**, 75 (2012).

[22] P. Jarillo-Herrero, J. A. van Dam, and L. P. Kouwenhoven, Nature **439**, 953 (2006).

[23] Jie Xiang, *et al.*, Nature-nanotechnology **1**, 208 (2006).

[24] M. Rosenau de Costa, I. A. Shelykh, and N. T. Bagraev, Phys. Rev. B **76**, 201302 (R) (2007).

[25] Chuan-Kui Wang and K.-F. Berggren, Phys. Rev. B **54**, R14257 (1996).

[26] A. A. Starikov, I. I. Yakimenko, and K.-F. Berggren, Phys. Rev. B **67,** 235319 (2003).

[27] R. Landauer, IBM J. Res. Dev., **1**, 233 (1957).

[28] M. Büttiker, Phys. Rev. Lett., **57**, 1761 (1986).

[29] V. J. Goldman, Phys. Rev. B **75**, 045334 (2007).

**Captions**

Fig. 1. Device schematic, showing the perspective view of the silicon sandwich structure performed within the framework of the Hall geometry. The silicon sandwich represents the p-type silicon quantum well confined by the δ-barriers heavily doped with boron on the n-type Si (100) surface. The source and drain Ohmic contacts are marked by S and D, respectively. The changes of the longitudinal voltage, $U_{xx}$, are measured by biasing the voltage applied to the Hall contacts, $V_{xy}$, when the drain-source current, $I_{ds}$, is stabilized at the extremely low value.

Fig. 2. Conductance measured at the temperature of 77K by biasing the voltage applied to the Hall contacts, $V_{xy}$, when the drain -source current was stabilized at the value of 0.5 nA. (a) Conductance increases as a function of $V_{xy}$ to a maximum of $4e^2/h$ demonstrating the standard plateaus, $2e^2/h$ and $3e^2/h$. The inset shows the conductance feature at the value of $15/4(e^2/h)$ corresponding to the middle of the step between the plateaus measured by high resolved biasing

the $V_{xy}$ voltage. (b) Fractional quantum conductance staircase close to the standard plateau at the value of $2e^2/h$. The dashed horizontal lines indicate the different fractional values. Insert shows the conductance plateaus and steps corresponding to the odd and even fractions. (c) Conductance measured as a function of $V_{xy}$ in the range of the values corresponding to the $0.7(2e^2/h)$ feature.

Fig. 3. Schematic diagram of the Si-QW confined by the δ–barriers heavily doped with boron, with the quantum point contact inserted as a result of the local disorder in the crystallographically-ordered chains of dipole centers of boron. The scheme used implies a vertical position of helical channels.

Fig. 4. Schematic of the spin-polarized edge channels in a quantum spin Hall insulator. Experimental setup on a six-terminal Hall bar showing pairs of edge states, with spin-up states in blue and spin-down states in red. Spin-dependent scattering near the Hall contact is provided by the varying the $V_{xy}$ voltage thus giving rise to the changes in the spin polarization degree and corresponding increase of conductance.

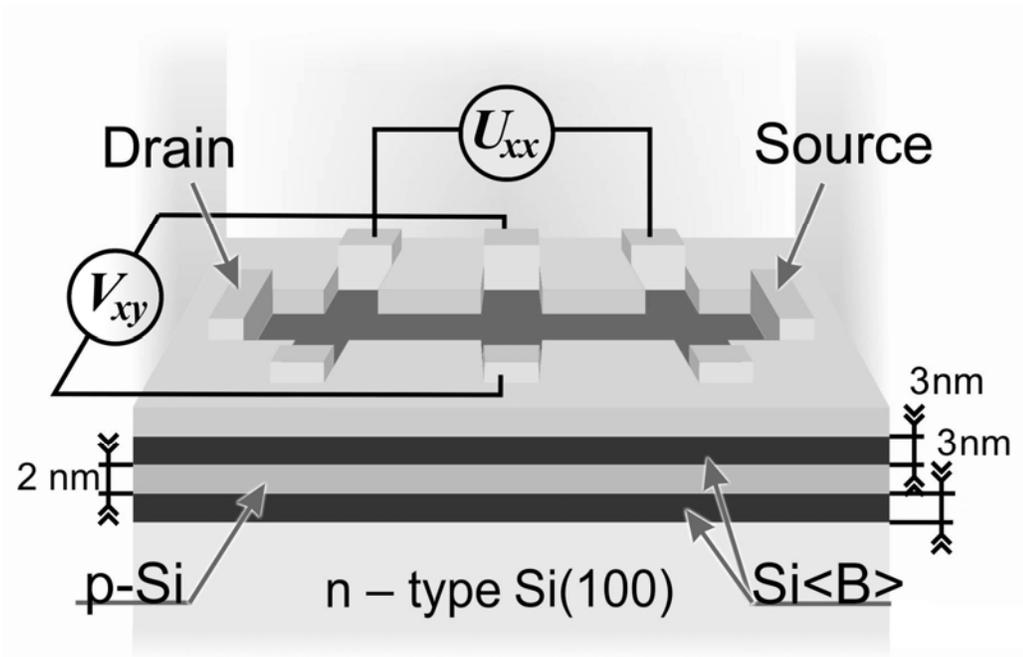

Figure 1.

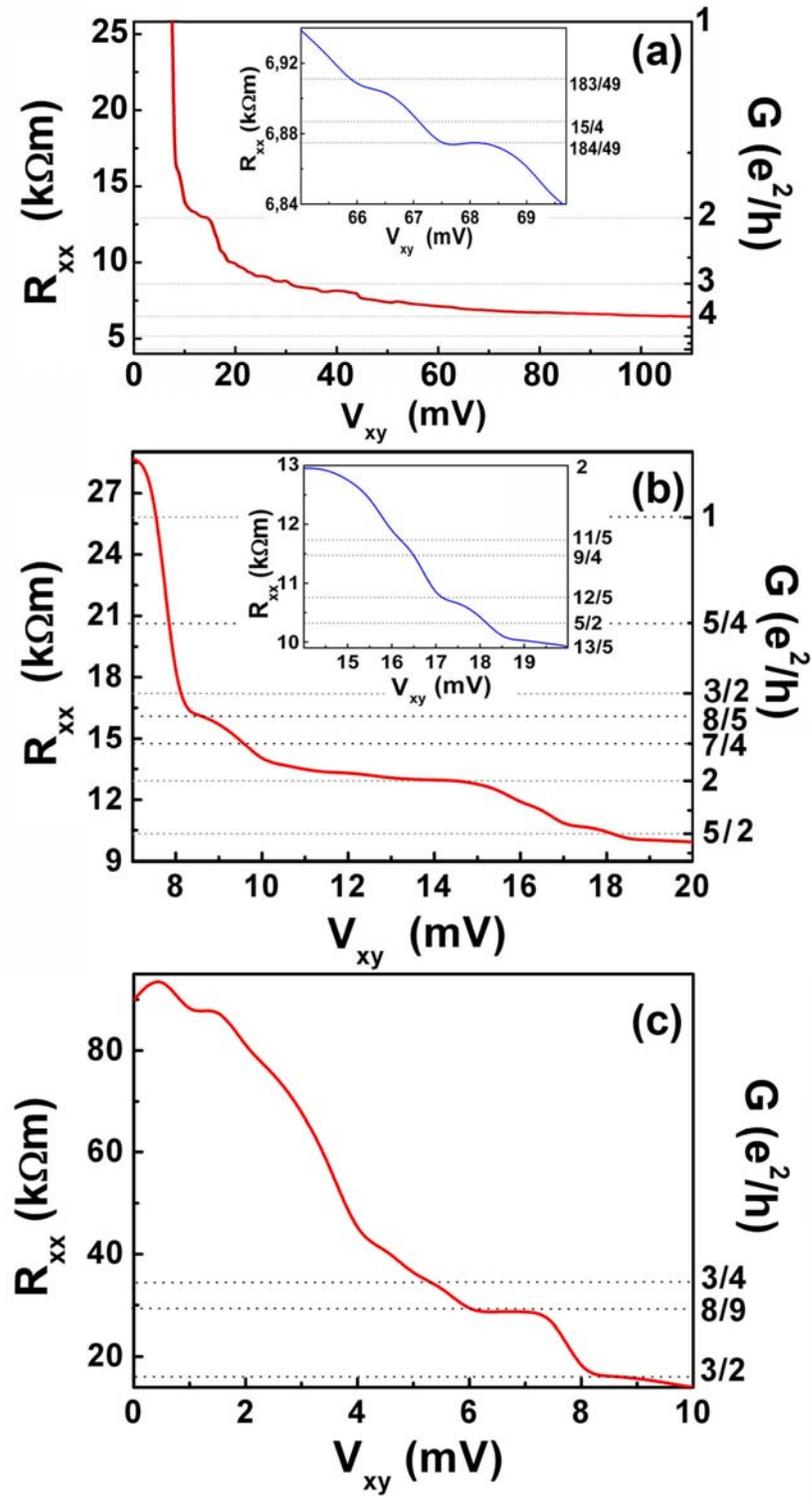

Figure 2.

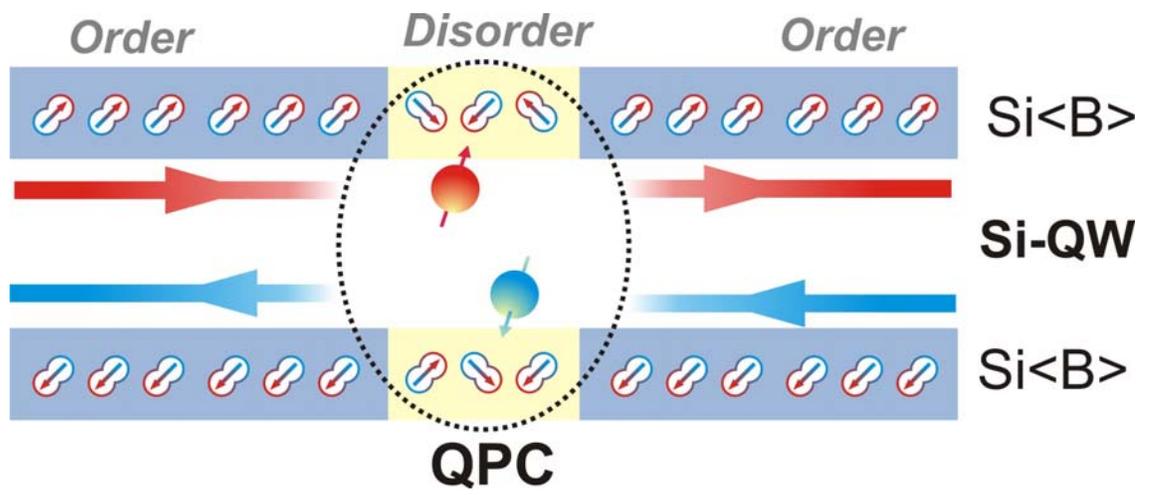

Figure 3.

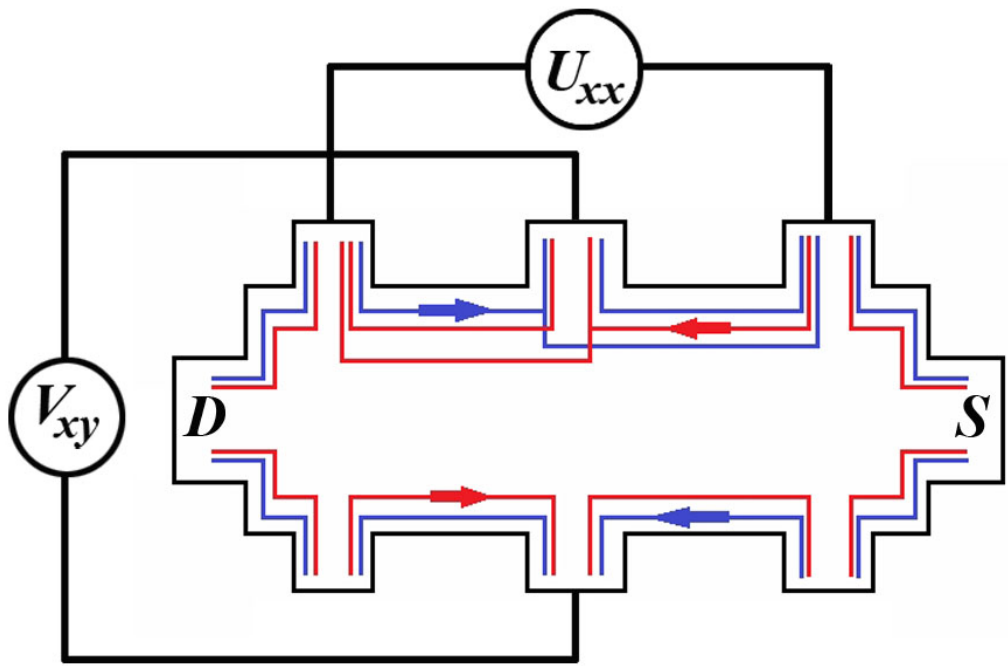

Figure 4.